\documentclass[prl,aps,twocolumn,amsmath,amssymb,superscriptaddress,nofootinbib]{revtex4-2}
\usepackage{url}
\usepackage{epsfig}
\usepackage{amsfonts}
\usepackage{graphicx}
\usepackage{epsfig}
\usepackage{eepic}
\usepackage{color}
\usepackage{bbm}
\usepackage{dcolumn}
\usepackage{bm}
\usepackage{ulem}
\usepackage{mathrsfs}
\usepackage{bbold}
\usepackage{datetime}
\usepackage{overpic} 
\usepackage{rotating}
\usepackage[usenames,dvipsnames]{xcolor}
\usepackage[colorlinks=true,citecolor=Magenta,linkcolor=Green,urlcolor=Blue]{hyperref}

\def\bc{\begin{center}}

\def\ec{\end{center}}
\def\be{\begin{eqnarray}}
\def\ee{\end{eqnarray}}
\definecolor{dyellow}{rgb}{1.,0.8,.0}
\definecolor{myblue}{rgb}{.1,.1,.7}
\definecolor{dcyan}{rgb}{.0,.6,.6}
\definecolor{dmagenta}{rgb}{0.6,0.0,0.6}
\definecolor{brown}{rgb}{0.6,0.2,0.}
\definecolor{darkblue}{rgb}{.0,.0,0.5}
\definecolor{darkred}{rgb}{0.75,0.0,0.0}
\definecolor{orange}{rgb}{1.,.6,.0}
\definecolor{dorange}{rgb}{0.8,.4,.0}
\definecolor{darkgreen}{rgb}{0.0,0.6,0.0}
\definecolor{purple}{rgb}{.4,.0,.4}
\definecolor{lightgrey}{rgb}{0.7, 0.7, 0.7}
\definecolor{grey}{rgb}{0.4, 0.4, 0.4}

\newcommand{\pap}[1]{\left( #1 \right)}

\begin{document}
\title{Universal Critical Holography and Domain Wall Formation}

\author{Tian-Chi Ma}\email{tianchima@buaa.edu.cn}
\affiliation{Center for Gravitational Physics, Department of Space Science, Beihang University, Beijing 100191, China}
\author{Han-Qing Shi}\email{by2030104@buaa.edu.cn}
\affiliation{Center for Gravitational Physics, Department of Space Science, Beihang University, Beijing 100191, China}
\author{Hai-Qing Zhang\href{https://orcid.org/0000-0003-4941-7432}{\includegraphics[scale=0.05]{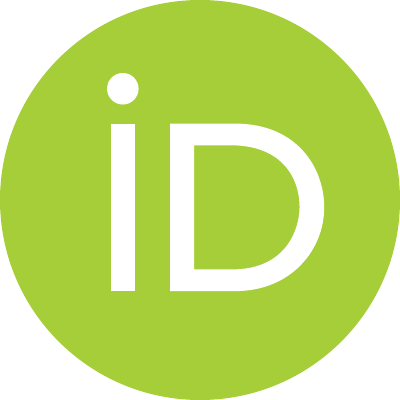}}}\email{hqzhang@buaa.edu.cn}
\affiliation{Center for Gravitational Physics, Department of Space Science, Beihang University, Beijing 100191, China}
\affiliation{Peng Huanwu Collaborative Center for Research and Education, Beihang University, Beijing 100191, China}
\author{Adolfo del Campo\href{https://orcid.org/0000-0003-2219-2851}{\includegraphics[scale=0.05]{orcidid.pdf}}}
\email{adolfo.delcampo@uni.lu}
\affiliation{Department  of  Physics  and  Materials  Science,  University  of  Luxembourg,  L-1511  Luxembourg, Luxembourg}
\affiliation{Donostia International Physics Center,  E-20018 San Sebasti\'an, Spain}

\begin{abstract}
Using holography, we study the universal scaling laws governing the coarsening dynamics of strongly coupled domain walls. Specifically, we studied the universal dependence of the length of the domain wall interfaces on the quench rate. The relation satisfies the Kibble-Zurek scaling shortly after the critical point. However, as time goes by, the coarsening dynamics suppresses the  Kibble-Zurek scaling in favor of a universal dynamical scaling of the characteristic length and the adiabatic growth of the system.   Theoretical predictions of the universal scaling laws are consistent with numerical findings in both regimes for both weak and strongly coupled systems.
\end{abstract}

\maketitle

\clearpage


A complete understanding of nonequilibrium phenomena involving many-body systems is currently lacking. This is particularly the case in systems far from equilibrium and at strong coupling. The Kibble-Zurek mechanism (KZM) is one of the few universal paradigms in this context \cite{Kibble76a,Kibble76b,Zurek96a,Zurek96b,Zurek96c}. It uses equilibrium scaling theory to describe the dynamics across a continuous or quantum phase transition. In particular, it exploits the scaling laws for the correlation length and the relaxation time,
\begin{equation}
\xi(\varepsilon)=\frac{\xi_0}{|\varepsilon|^{\nu}},  \quad\tau(\varepsilon)=\frac{\tau_0}{|\varepsilon|^{z\nu}}.
\end{equation}
Here, $\nu$ and $z$ are critical exponents and $\xi_0$ and $\tau_0$ are system-dependent microscopic constants, while $\varepsilon=(\lambda-\lambda_c)/\lambda_c$ determines the proximity of the control parameter $\lambda$ to its critical value $\lambda_c$.
The divergence of the relaxation time in the neighborhood of the critical point, known as the 
critical slowing down, plays a key role in the KZM. Consider driving the system by a linear modulation of the control parameter $\lambda$, such that $\varepsilon=t/\tau_Q$, where $\tau_Q$ is the quench time, and the critical point is reached at $t=0$. 
The timescale in which the instantaneous equilibrium relaxation time matches the time elapsed after crossing the critical point is known as the freeze-out time and plays a key role in KZM. It is predicted to scale universally with the quench time according to 
\begin{equation}
    \hat{t}=\left(\tau_0\tau_Q^{\nu z}\right)^{\frac{1}{1+\nu z}}.
    \label{t-hat}
\end{equation}
Using it, the KZM predicts the nonequilibrium correlation length to be given by
\begin{equation}
\hat{\xi}=\xi[\varepsilon(\hat{t})]=\xi_0   \left(\frac{\tau_{Q}}{\tau_0}\right)^{\frac{ \nu}{1+z\nu}}.
\label{KZMxi}
\end{equation}
It follows that the density of topological defects of dimension $d$ in a system in a $D$ dimensional space is given by.
\begin{equation}
 \rho= \frac{1}{\hat{\xi}^{D-d}}=\frac{1}{\xi_0^{D-d}}\left(\frac{\tau_0}{\tau_Q}\right)^{\frac{(D-d)\nu}{1+z\nu}}.
 \label{rhoKZMeq}
\end{equation}

A large body of accumulated evidence supports the KZM, relying on exact solutions, numerical simulations, and experimental studies \cite{Dziarmaga10,DZ14}.
In particular, the last decade has witnessed remarkable experimental progress thanks to advances in quantum simulation in platforms including trapped ions \cite{Ulm13,Pyka13,Cui16,Cui20} and Rydberg gases \cite{Lukin17}, the development of programmable quantum annealing devices \cite{Weinberg20,Bando20,King22}, and the improved probing techniques in condensed matter experiments \cite{Lin14,Du23}. In addition, universal physics beyond the scope of KZM has been discovered, governing the defect number statistics \cite{delcampo18,Cui20,GomezRuiz20,Bando20,Mayo21,delcampo21,King22} and the spatial distribution of defects \cite{delcampo22,thudiyangal2024universal}. 

Despite this progress, KZM and its generalizations have been mostly studied in scenarios characterized by point-like topological defects with $d=0$, such as parity-breaking kinks   \cite{Laguna98,Sabbatini11,Ulm13,Pyka13,Krapivsky10} and solitons \cite{Damski10} in $D=1$,  as well as vortices in $D=2$  \cite{Weiler08,Griffin12,Lin14,delcampo21,Goo21,delcampo22,Rabga23,thudiyangal2024universal}.  Evidence supporting KZM for extended topological defects with $d\geq 1$ remains limited. In such a setting, deviations may arise due to the enhancement of coarsening \cite{Libal20,Samajdar24}, the role of the embedding geometry \cite{Stoop18}, and effects arising from the spatial configuration of the extended defect, with no counterpart in the point-like case. The latter can stem from conformational entropy, intra-defect interactions, and excitations of modes propagating along the extended degrees of freedom, such as Kelvin modes in vortex strings. 
Numerical simulations have described 3D $U(1)$ vortex string formation following KZM scaling laws \cite{Antunes99}, while recent simulations of skyrmions are better explained by coarsening \cite{Reichhardt23}. Experiments on 
Rayleigh-B\'enard convection with convective rolls and cells as defects match the KZM prediction \cite{Casado06}.  
Domain formation on colloidal monolayers, probing the Kosterlitz–Thouless–Halperin–Nelson–Young (KTHNY) melting scenario, exhibits important deviations from the conventional KZM due to the nature of the transition and suggests unusually large values of the dynamic critical exponent \cite{Keim15,delcampo15}. By contrast,  recent experiments on domain wall formation of 3D Ising systems are consistent with the KZM scaling  \cite{Du23}; see as well \cite{Dziarmaga04} for supporting numerics in 2D. 

Coarsening is governed by a dynamic critical exponent $z_d$, which is generally different from the equilibrium value $z$. According to Biroli et al. \cite{Biroli10}, for $t\gg\hat{t}$, the nonequilibrium correlation length scales as 
\begin{equation}
R(t)=\xi[ \varepsilon(t)]^{1-\frac{z}{z_d}}t^{\frac{1}{z_d}}.    
\end{equation}
where $\xi[ \varepsilon(t)]=\varepsilon(t)^{-\nu}=(t/\tau_Q)^{-\nu}$. 

In this study, we investigate the formation of domain walls in a holographic setting \cite{Hartnoll18book}. Holography provides a framework to explore strongly-coupled field theories, in and out of equilibrium, and has been used to explore the validity of KZM \cite{Chesler15,Sonner15,delcampo21,Xia23}. 
We focus on recent developments allowing for the description parity symmetry breaking, which results in domain formation analogous to that in a ferromagnet. In this setting, we establish the universality of the critical dynamics by characterizing the domain wall average length and its statistics beyond the scope of KZM, revealing a crossover from KZM to coarsening-dominated scaling laws.

\begin{figure}[t]
\centering
\includegraphics[trim=7cm 3cm 4.8cm 3cm, clip=true, scale=0.3]{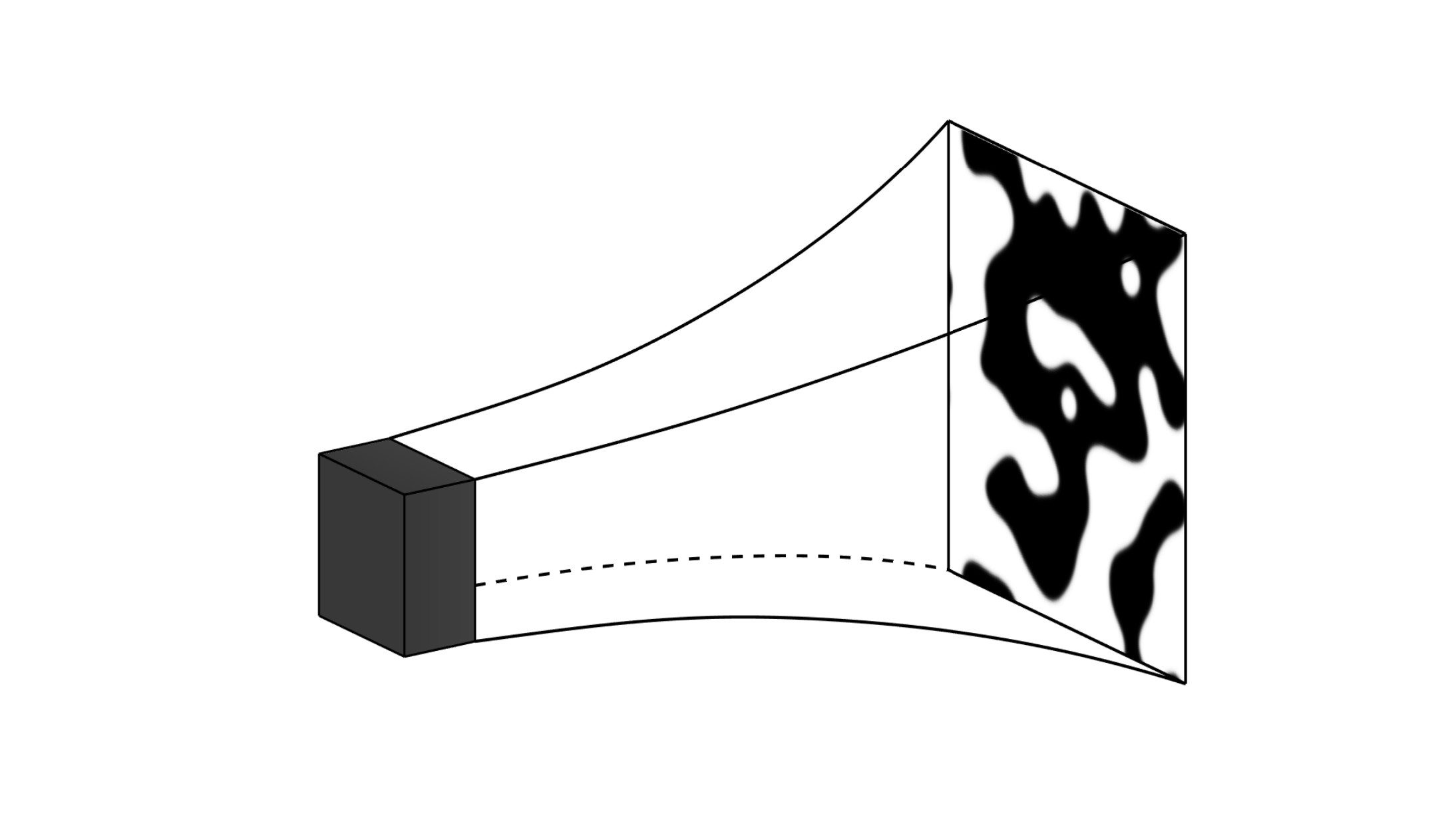}
\put(-55,130){\begin{rotate}{-24}\bf\text{Boundary}\end{rotate}}
\put(-225,70){\begin{rotate}{-0}\bf\text{Black Hole}\end{rotate}}
\put(-148,44){\begin{rotate}{-0}\bf\text{AdS$_4$\,:\,($\Psi$,\,$M_\mu$)}\end{rotate}}
\caption{Schematic figure of the holographic mapping.}\label{3D} 
\end{figure}

{\it Background of gravity. }
The schematic setting to probe the holographic domain wall formation is shown in Fig. \ref{3D}. The gravity background we choose is the AdS$_4$ black brane in Eddington-Finkelstein coordinates,
 \be \label{metric}
 ds^2=\frac{1}{z^2}\left[-f(z)dt^2-2dtdz+dx^2+dy^2\right],
 \ee
 in which $f(z)=1-(z/z_h)^3$ with $z_h$ the horizon position. The AdS infinite boundary is at $z = 0$ where the field theory lives. The Lagrangian of the model we adopt is the usual Abelian-Higgs model for holographic superconductors,
 \be\label{action}
 \mathcal{L}=-\frac14F_{\mu\nu}F^{\mu\nu}-|D_\mu\tilde\Psi|^2-m^2|\tilde\Psi|^2,
 \ee
 where $F_{\mu\nu}=\partial_\mu A_\nu-\partial_\nu A_\mu$ and $A_\mu$ is the $U(1)$ gauge field,  $D_\mu=\nabla_\mu-iA_\mu$, and $\tilde\Psi$ is the complex scalar field.  We work in the probe limit, in which the equations of motion (EoMs) read
 \be\label{eq1}
 &&D_\mu D^\mu \tilde\Psi-m^2\tilde\Psi=0,\nonumber \\  &&\nabla_\mu F^{\mu\nu}=i\left(\tilde\Psi^*D^\nu\tilde\Psi-\tilde\Psi(D^\nu\tilde\Psi)^*\right).
 \ee
 To transform the Lagrangian with $U(1)$ symmetry to $Z_2$ symmetry, we must transform the complex scalar fields into real ones. To this end, we make the following transformations \cite{Horowitz:2011dz,Li:2022tab}
 \be \tilde\Psi=\Psi e^{i\lambda},\qquad  A_\mu=M_\mu+\partial_\mu\lambda, \label{gaugefix}\ee
 where $\Psi, M_\mu$ and $\lambda$ are real functions. With these real functions, the previous Eq. \eqref{eq1} can be rewritten as 
 \be\label{eq2}
 \mathbb{D}_\mu\mathbb{D}^\mu\Psi-m^2\Psi=0, \qquad  \nabla_\mu F^{\mu\nu}=2M^\nu\Psi^2, 
 \ee
 where $\mathbb{D}_\mu=\nabla_\mu-iM_\mu$. We use the following  ans\"atze for the fields: $\Psi=\Psi(t,z,x,y)$, $M_t=M_t(t,z,x,y)$, $M_z=M_z(t,z,x,y)$, $M_x=M_x(t,z,x,y)$, and  $M_y=M_y(t,z,x,y)$. The details of the EoMs can be found in \cite{SM}.

\begin{figure*}[ht]
	\includegraphics[trim=0cm 0cm 0cm 0cm, clip=False, scale=0.47]{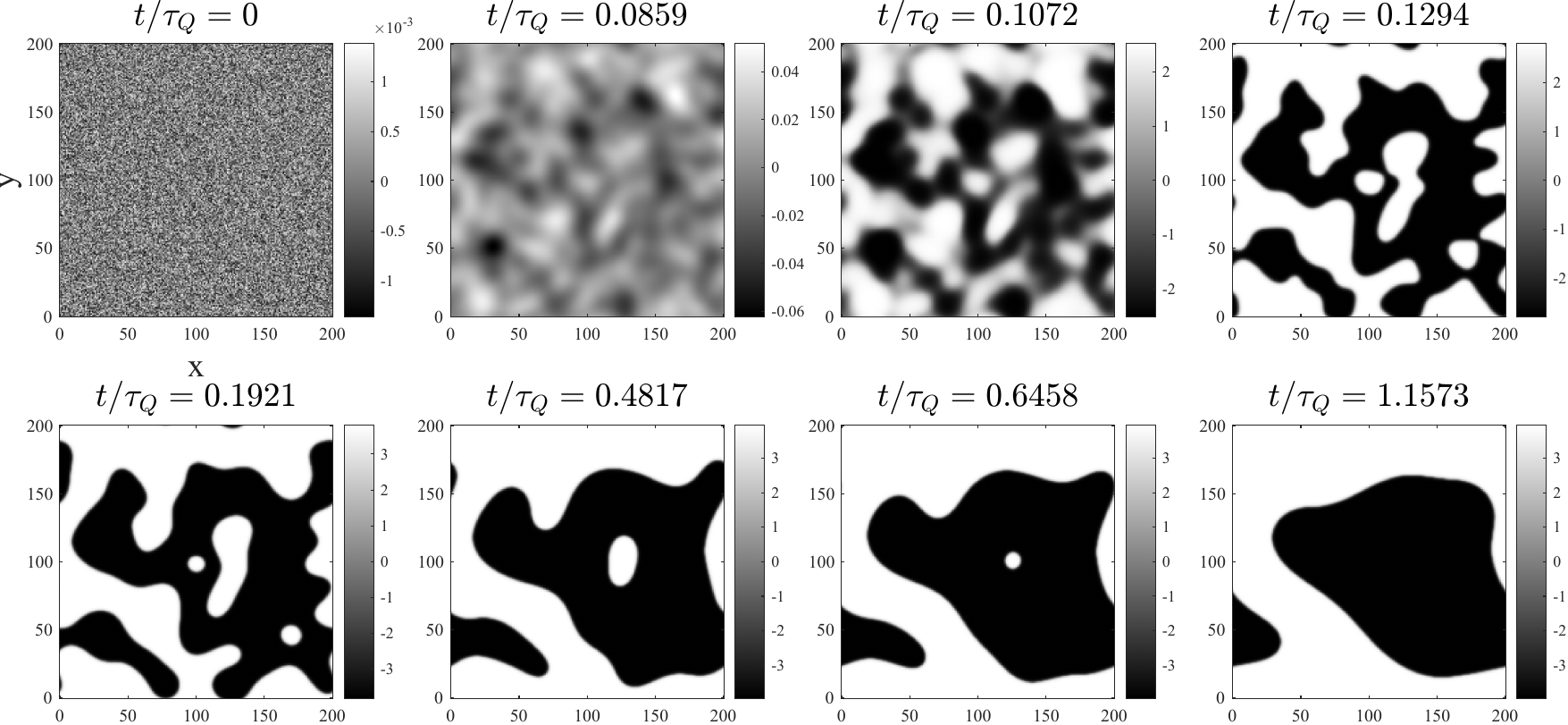}
	\includegraphics[trim=12cm 0cm 0cm 0cm, clip=False, scale=0.54,angle=-90]{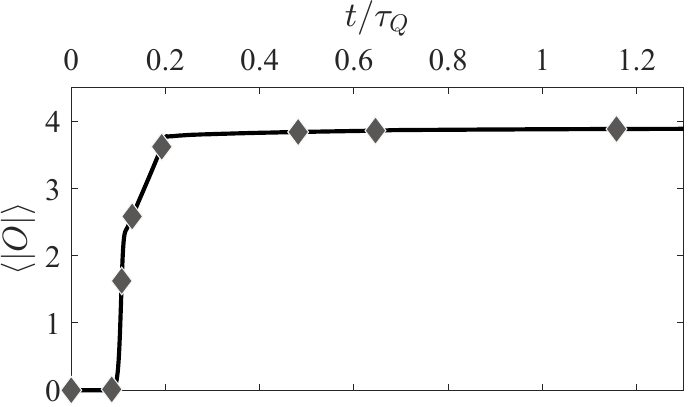}
	\put(-300,-10){\small\bf(a)}
	\put(-65,-10){\small\bf(b)}
    \caption{ Time evolution of the order parameter and the birth of domain wall in the AdS boundary. (a) Density plots of the evolving order parameter at eight specific times 
    with $\tau_Q$ = $e^{6.25}$. One can see how the order parameters evolve from initial random configurations into domain structures. (b) The average absolute value of the order parameter $\langle |O|\rangle$ during quenches.  Each diamond corresponds to one of the snapshots in panel (a). }\label{hoevolution}
\end{figure*}

{\it Boundary conditions.}   The asymptotic behavior of fields near $z\rightarrow0$ is described by $\Psi\sim{z}(\Psi_0(t,x,y)+\Psi_1(t,x,y)z+...)$, $M_t\sim\mu(t,x,y)-\rho(t,x,y)z+...$, $M_z\sim a_z(t,x,y)+b_z(t,x,y)z+...$, $M_x\sim a_x(t,x,y)+b_x(t,x,y)z+...$, $M_y\sim a_y(t,x,y)+b_y(t,x,y)z+...$. We
have set the scalar field mass square as $m^2=-2$ and a unit AdS radius. By choosing the standard quantization setting $\Psi_1\equiv0$, it follows that $\Psi_2$ is related to the condensate of the superconducting order parameter $O(t,x,y)$ in the boundary field theory. The parameters $\mu$ and $\rho$ are interpreted as the chemical potential and charge density, respectively, at the
boundary. At the horizon, we set $M_t(z_h)=0$ and the regular finite boundary conditions for other fields. In addition, we use the periodic boundary conditions for all the fields along $(x,y)-$directions. 
The fourth-order Runge-Kutta method is used to propagate the system with time step $\triangle t=0.01$. In the radial direction $z$, the Chebyshev pseudo-spectral method with $21$ grid points is used. In the $(x,y)$-direction,  we set the length of the space as  $L_x=L_y=200$, and we use the Fourier decomposition with $201\times201$ grid points. We add the random seeds of the fields in the bulk by satisfying the distributions $\langle s(t,x_i)\rangle=0$ and $\langle s(t,x_i)s(t',x_j)\rangle=h\delta(t-t')\delta(x_i-x_j)$, with the amplitude  $h=10^{-3}$. 

{\it Cooling the system.} 
In the setup of a holographic superconductor, decreasing $T$ is equivalent to increasing $\mu$. To linearly quench the temperature across the critical point as $T(t)/T_c=1-t/\tau_Q$, we vary the chemical potential as
\be\label{muquench}
\mu(t)=\mu_c/(1-t/\tau_Q),
\ee 
where $\tau_Q$ is the quench rate and  $\mu_c=4.06$ denotes the critical potential in the static case. We drive the system from the initial temperature $T_i=T_c$ to the superconducting state with the final equilibrium temperature $T_f=0.8T_c$. 

Following a thermal quench from high to low temperatures across the critical point $T_c$, the system enters the superconducting phase, and domain walls appear as predicted by the KZM. Figure \ref{hoevolution} shows the evolution of the order parameter in panel (a)  and its average absolute value in panel (b) during the transition from a homogeneous phase to the symmetry-breaking phase. Panel (a) shows eight snapshots at different times of evolution  
for a fixed quench time $\tau_Q=e^{6.25}$, 
each of which is marked by a diamond symbol in pane (b).
At the initial time $t/\tau_Q$ = 0, the order parameter exhibits a random configuration similar to a mosaic picture and then evolves and grows rapidly around $t/\tau_Q$ = 0.0859. At the time $t/\tau_Q$ = 0.1072, the system is divided into domains, but the order parameter still exhibits smooth gradients. From time $t/\tau_Q$ = 0.1294 to  $t/\tau_Q$ = 1.1573, the average value of $\langle |O|\rangle$ increases linearly and saturates at a plateau. As the coarsening dynamics becomes more prominent, it reduces the small-scale structures in the nonequilibrium configuration, leading to the coalescence of domains into bigger ones. 

{\it Scaling of the average domain-wall length.}
In the domain formation associated with the breaking of parity in two spatial dimensions, the number of defects can be associated with the total domain wall length, which we denote by $L$, averaged over an ensemble of realizations.
Adapting the KZM scaling law in Eq. (\ref{rhoKZMeq}), the generic universal dependence on the quench time $\tau_Q$ reduces to 
\be
\label{scaling relation}
L \propto \tau_{Q}^{-\left(D-d\right)\nu/\left( 1+z\nu\right)}\propto \tau_{Q}^{-\nu/\left( 1+z\nu\right)},
\ee 
in two spatial dimensions ($D=2$) for domain walls ($d=1$)  in the AdS boundary.  From AdS/CFT, the boundary field theory is a mean-field theory. As a result,  $\nu=1/2$ and $z=2$, and thus $L \propto \tau_{Q}^{-1/4}$. 

\begin{figure}[t]
	\centering
	\includegraphics[trim=0cm 0cm 0cm 0cm, clip=true, scale=0.31]{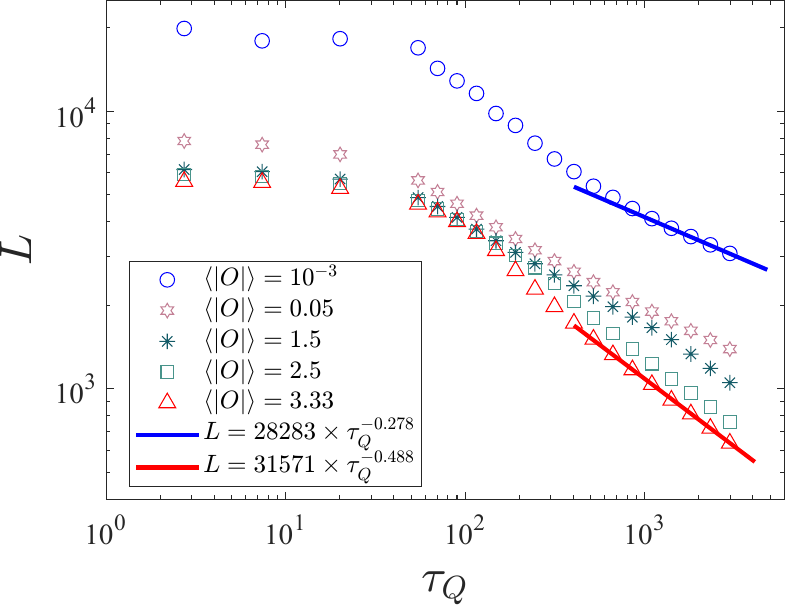} 
	\put(-115,95){(a)}~
	\includegraphics[trim=0cm 0cm 0cm 0cm, clip=true, scale=0.31]{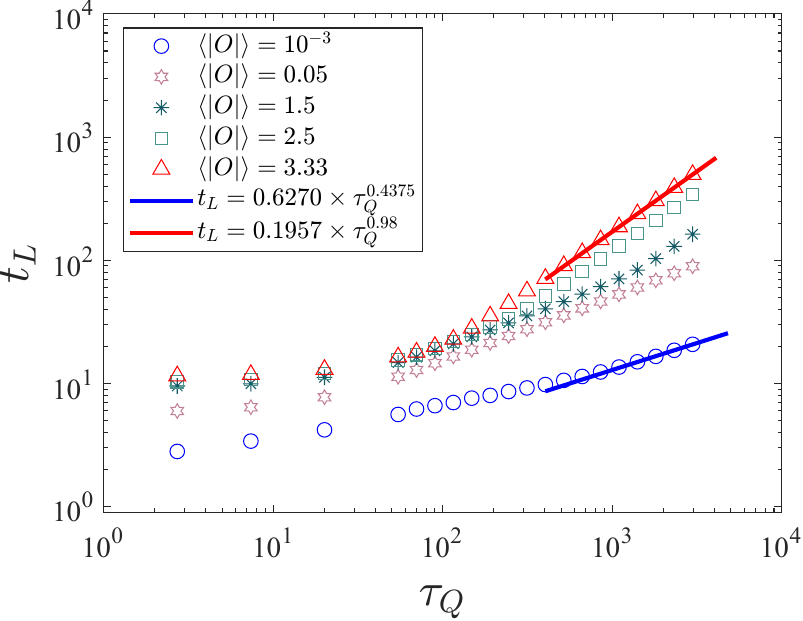}
	\put(-115,95){(b)}
	\caption{(a) Dependence of the domain wall length $L$ on the quench rate $\tau_Q$ for the average absolute value of the order parameter $\langle |O|\rangle$. 
 (b) Lag time as a function of the quench rate for various condensates. The circles, hexagons, stars, squares, and triangles are the numerical data, while the straight lines are the best-fitted lines in the slow quench regime.
 }
 \label{hoquench} 
\end{figure}

Figure \ref{hoquench}(a) shows the relations between domain wall length $L$ and the quench rate $\tau_Q$ for several values of the average absolute value of the order parameter $\langle |O|\rangle$, using   2000 independent trajectories.  Figure  \ref{hoquench}(a) indicates that for different $\langle |O|\rangle$, there exist different scaling relations between $L$ and $\tau_Q$. Specifically, as $\langle |O|\rangle=10^{-3}$, the fitted line has a power-law scaling  $L = 28283\times\tau_{Q}^{-0.278}$, which is close to the KZM prediction $-\left(D-d\right)\nu/\left( 1+z\nu\right) = -1/4$ in Eq. (\ref{scaling relation}). However, when $\langle |O|\rangle=3.33$, the fitted line is $L = 31571\times\tau_{Q}^{-0.488}$, and the scaling deviates from the theoretical KZM prediction. This behavior is attributed to coarsening, as it takes more time for a slowly quenched system to achieve the same value of $\langle |O|\rangle$ than for a rapidly quenched system.  The coarsening process always exists after the formation of defects. Therefore, to achieve the same $\langle |O|\rangle$ value, a slowly quenched system must undergo a longer coarsening evolution than a rapidly quenched system. As a result, coarsening competes with the KZM scaling.

According to KZM, the instant at which the system ``freezes out'' is set by $\hat t$ in Eq. (\ref{t-hat}). In practice, one defines the lag-time $t_L$ by the response of the system to represent the freeze-out time \cite{Das12,Chesler15,Sonner15,thudiyangal2024universal}. In theory, where $t_L\propto \tau_Q^{1/2}$ in the holographic setup. Figure \ref{hoquench}(b)  exhibits the relation between the lag time and the quench rate for various condensates. When the condensate is small, $\langle |O|\rangle\approx 10^{-3}$, the scaling exponent is roughly $0.4375$, which is close to the KZM prediction. By contrast, for large condensates (e.g., when $\langle |O|\rangle\approx3.33$), the scaling is roughly $0.98$, which rules out the KZM prediction. 
This is an important advantage of our setting over previous one-dimensional studies of parity-symmetry breaking in which the power-law exponent associated with KZM \cite{Laguna98,DeChiara2010} and coarsening \cite{Krapivsky10,Mayo21} happen to coincide, hindering the study of their interplay.
We expect that the deviations of the scaling in the $t_L\sim\tau_Q$ relation are also due to the coarsening in the late stage of the evolution.

\begin{figure}[t]
	\centering
	\includegraphics[trim=0cm 0cm 0cm 0cm, clip=true, scale=0.31]{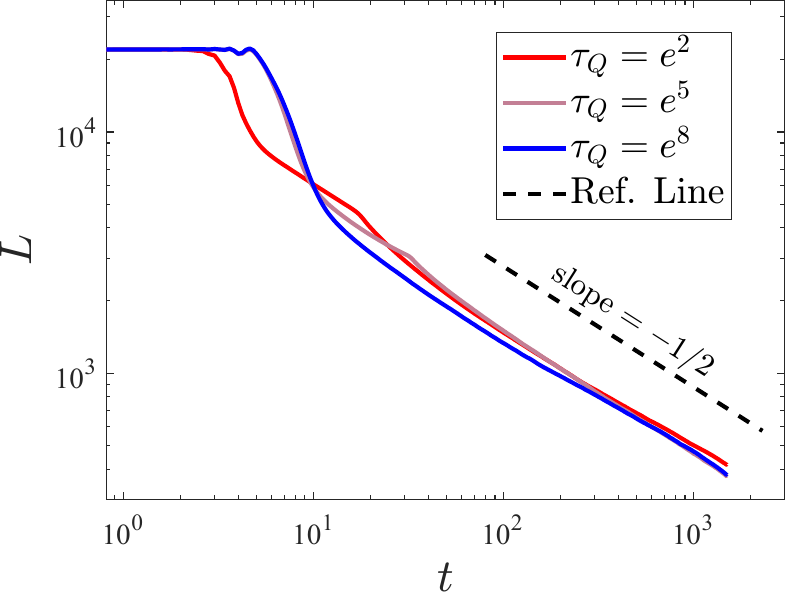}
 \put(-115,95){(a)}~
 \includegraphics[trim=0cm 0cm 0cm 0cm, clip=true, scale=0.31]{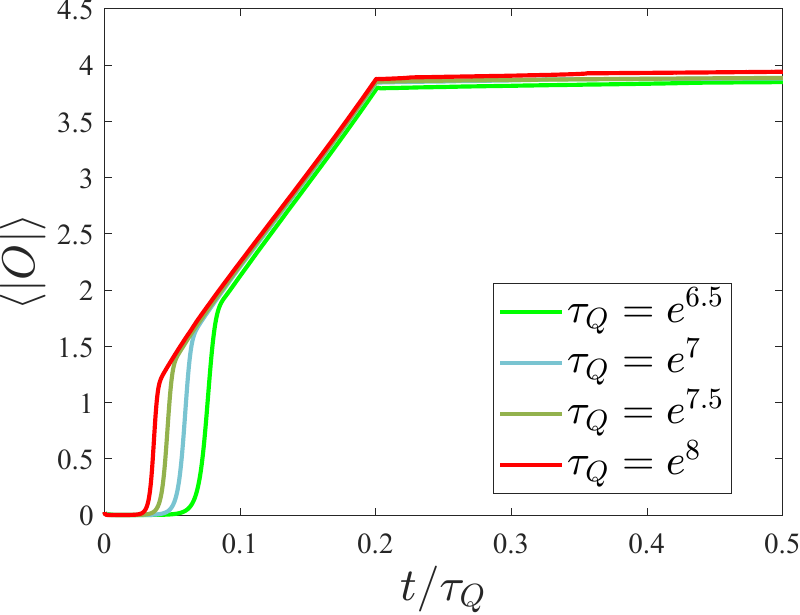}
 \put(-115,95){(b)}
	\caption{(a) Time evolution of domain wall length for various quench rate $\tau_Q$.  
 The dashed line is the theoretical reference line with a slope of $-1/2$; (b) Average condensate for slow quenches. The linear relation reflects the adiabatic growth. }\label{lengthofDW} 
\end{figure}

{\it Domain coarsening.}
Coarsening involves gradient flow dynamics. The relaxation of a non-equilibrium system follows the steepest descent in the energy landscape.
Adapting the prediction by Biroli et al. \cite{Biroli10,Samajdar24}, the domain wall length scales as 
\begin{equation}\label{LR}
L\propto \frac{1}{R(t)^{D-d}}\propto\tau_Q^{\frac{(D-d)\nu}{z_d}(z-z_d)}t^{-\frac{(D-d)}{z_d}[1+\nu(z-z_d)]}.    
\end{equation}
For $z=z_d=2$, $L\propto t^{-\frac{(D-d)}{z_d}}\propto t^{-1/2}$  and $R(t)\propto t^{1/2}$, which is in agreement with the results in \cite{Bray:1994zz}.
In Fig. \ref{lengthofDW}(a), we present the length of domain walls $L$ as a function of $t$ for several values of $\tau_Q$. Long after the quench,  it can be found that regardless of the value of $\tau_Q$, the time evolution of $L$ follows a similar power law and is almost parallel, in log-log scale,  to the reference line $L\sim t^{-1/2}$,  which agrees with the theoretical prediction in Eq. \eqref{LR}. At this late stage, the system has reached a state of local equilibrium, and coarsening dynamics completely govern the evolution of the system. 

Fig. \ref{lengthofDW}(b) shows that the condensates transit from exponential growth to a common linear growth as a function of $t/\tau_Q$, for large $\tau_Q$. This is an implication of the adiabatic evolution \cite{Chesler15}. For large values of $\tau_Q$, the condensate scales with the same function of $t/\tau_Q$. In this regime,   when $\langle |O|\rangle$ is large (e.g., $\langle |O|\rangle=3.33$)  the lag-time is linearly proportional to $\tau_Q$. This is consistent with the numerical relation in Fig. \ref{hoquench}(b) that $t_L\propto \tau_Q^{0.98}$ as $\tau_Q$ is large. From the relations $L\sim t^{-1/2}$ and $t\sim\tau_Q$ after the coarsening dynamics, we deduce that  $L\sim\tau_Q^{-1/2}$, which is consistent with the relation between $L$ and $\tau_Q$ in Fig. \ref{hoquench}(a) as $\langle |O|\rangle=3.33$.

\begin{figure}[t]
	\centering
	\includegraphics[trim=0cm 0cm 0cm 0cm, clip=true, scale=0.32]{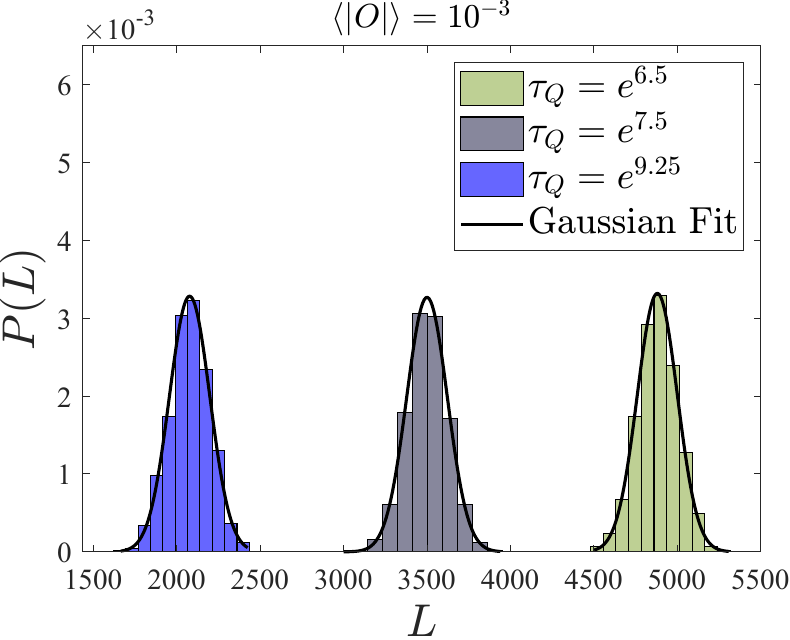}\put(-115,100){(a)}
	\includegraphics[trim=0cm 0cm 0cm 0cm, clip=true, scale=0.32]{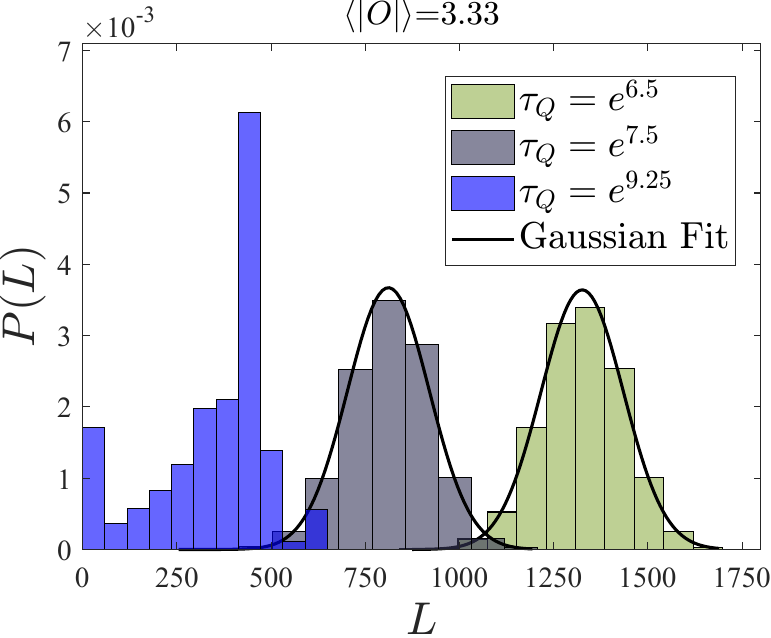}\put(-115,100){(b)}
    \caption{(a) Histogram of the probability of the domain wall length, for different values of $\tau_Q$ at $\langle|O|\rangle=10^{-3}$, are well fitted by a Gaussian distribution. (b)  Histogram of the domain wall length probability for different values of $\tau_Q$ at $\langle|O|\rangle=3.33$. Coarsening-induced deviations from the Gaussian fit arise in the slow quench limit, leading to a multimodal distribution.
    }\label{histo} 
\end{figure}
Additional information is revealed by the domain wall statistics.
Fig. \ref{histo}(a) shows the histogram of the probability of the domain wall lengths for three kinds of quench rates, 
and a small condensate $\langle |O|\rangle=10^{-3}$. In this regime, the nonequilibrium dynamics is consistent with the KZM. The universality of average quantities with the quench rate carries over the full distribution, according to the recent extension beyond KZM \cite{delcampo18,GomezRuiz20}. Numerical evidence to date supporting such extensions is limited to point-like defects in one and two spatial dimensions \cite{delcampo18,Cui20,GomezRuiz20,Bando20,Mayo21,delcampo21,King22}. 
For large system sizes, the distribution approaches a normal distribution, justifying the accuracy of the Gaussian fit in Fig. \ref{histo}(a). This provides the first numerical evidence regarding the universal character of the distribution of domain walls as extended topological defects with $d=1$. Such universality remains in larger condensates as shown in Fig. \ref{histo}(b) for $\langle|O|\rangle=3.33$ for quenches $\tau_Q=e^{6.5}$ and $\tau_Q=e^{7.5}$. However, the distribution for slow quenches $\tau_Q=e^{9.25}$ is dominated by coarsening; it becomes bimodal and no longer universal, with increasing probability near the origin, with low defect numbers.

{\it Discussion and Conclusions.}
We have established the universal scaling laws governing the critical dynamics leading to strongly coupled domain walls via the AdS/CFT correspondence. The KZM describes the universal scaling of the domain wall length with the quench time near the critical regime. Far away from this regime, a different scaling holds that is no longer consistent with  KZM and reflects the coarsening dynamics of domain walls.  A theoretical analysis yields a universal prediction for the scaling dominated by coarsening that is in agreement with numerical simulations.  
In \cite{SM},
we further show the similar phenomenon of domain wall in the Ginzburg-Landau model, which is a weakly coupled theory. Therefore, we can conclude that our findings of the new scalings in the far-from critical regime may apply universally in the coarsening dynamics. 

{\it Note:} Upon the completion of this work, Ref. \cite{andersen2024thermalization} reported KZM deviations due to coarsening in the digital-analog quantum simulation of a 2D XY ferromagnet. 

{\it Acknowledgements.} 
This work was partially supported by the National Natural Science Foundation of China (Grants No. 12175008). This research was funded in part by the Luxembourg National Research Fund (FNR), grant No. 17132060. For open access, the authors have applied a Creative Commons Attribution 4.0 International (CC BY 4.0) license to any Author Accepted Manuscript version arising from this submission.

\bibliographystyle{apsrev4-2}
\bibliography{HoloDomainWallsBib}

\newpage
\widetext

\appendix

\begin{center}
\label{app}
\end{center}

\setcounter{equation}{0}
\setcounter{figure}{0}
\setcounter{table}{0}
\setcounter{section}{0}
\makeatletter
\renewcommand{\theequation}{S\arabic{equation}}
\renewcommand{\thefigure}{S\arabic{figure}}
\renewcommand{\bibnumfmt}[1]{[#1]}
\renewcommand{\citenumfont}[1]{#1}

\section{Appendix A: Explicit forms of the equations of motions}
\label{eomdetail}

The EoMs \eqref{eq2} in the main text can be decomposed into 
\be\label{eq_app}
\nabla_\mu\nabla^\mu\Psi-M_\mu M^\mu\Psi-m^2\Psi&=&0, \label{eqapp1}\\
\left(\nabla_\mu M^\mu\right)\Psi+2M^\mu\nabla_\mu \Psi&=&0,\label{eqapp2}\\
\nabla_\mu F^{\mu\nu}&=&2M^\nu\Psi^2. \label{eqapp3}
\ee
The Eqs. \eqref{eqapp1} and \eqref{eqapp2} follow respectively from the real and imaginary parts of the scalar equations in Eq.\eqref{eq2}. We note that Eqs. \eqref{eqapp2} and \eqref{eqapp3} are not independent, since 
\be \nabla_\nu(\nabla_\mu F^{\mu\nu})=0=2(\nabla_\nu M^\nu) \Psi^2+4M^\nu\Psi(\nabla_\nu\Psi). \label{cons}\ee
Dividing $2\Psi$ on the right side of the Eq. \eqref{cons}, we can readily get the left side of the Eq. \eqref{eqapp2}. Therefore, there are five real functions with five independent equations, which implies that our ans\"atze for the fields are consistent. Using them in the frame of the line-element \eqref{metric}, the EoMs involve the following contributions:  \\
1) The gauge fields \eqref{eqapp3} part,
\be
0&=&-\frac{2\Psi^2M_t}{z^2}+\partial^2_xM_t+\partial^2_yM_t+f\partial^2_zM_t-\partial_{tx}M_x-\partial_{ty}M_y 
-\partial_{tz}M_t 
-f\partial_{tz}M_z+\partial^2_tM_z,\label{em1}\\
0&=&-\frac{2\Psi^2M_z}{z^2}+\partial^2_xM_z+\partial^2_yM_z-\partial_{zx}M_x-\partial_{zy}M_y+\partial^2_zM_t-\partial_{tz}M_z,\label{em2}\\
0&=&-\frac{2\Psi^2M_x}{z^2}+\partial^2_yM_x-f'\partial_xM_z-\partial_{xy}M_y+f'\partial_zM_x+\partial_{zx}M_t-f\partial_{zx}M_z
+f\partial^2_zM_x+\partial_{tx}M_z-2\partial_{tz}M_x,\label{em3}\\
0&=&-\frac{2\Psi^2M_y}{z^2}+\partial^2_xM_y-f'\partial_yM_z-\partial_{xy}M_x+f'\partial_zM_y+\partial_{zy}M_t-f\partial_{zy}M_z 
+f\partial^2_zM_y+\partial_{ty}M_z-2\partial_{tz}M_y. \label{em4}
\ee
2) The real part of scalar fields \eqref{eqapp1}, 
\be
0&=&-\frac{m^2\Psi}{2z^2}-\frac12\Psi M_x^2-\frac12\Psi M_y^2+\Psi M_t M_z-\frac12\Psi f M_z^2+\frac12\partial^2_x\Psi+\frac12\partial^2_y\Psi \nonumber\\
&&-\frac{f\partial_z\Psi}{z}+\frac12\partial_z(f\partial_z\Psi)
+\frac{\partial_t\Psi}{z}-\partial_{tz}\Psi.
\ee
3) The imaginary part of scalar fields \eqref{eqapp2},
\be
0&=&\frac{2\Psi}{z}(fM_z-M_t)-\Psi\left(\partial_z(fM_z)+\partial_x M_x+\partial_y M_y-\partial_tM_z-\partial_zM_t\right) \nonumber\\
&&+2\left(M_t\partial_z\Psi+M_z\partial_t\Psi-M_x\partial_x\Psi-M_y\partial_y\Psi-fM_z\partial_z\Psi\right),
\ee
in which $f'=f'(z)$.

\section{Appendix B: Domain wall formations in a time-dependent Ginzburg-Landau model}
\label{GLmodel}
We start with a time-dependent Ginzburg-Landau model (TDGL) of the real non-conserved scalar field $\phi$  undergoing Langevin dynamics
\begin{equation}\label{GLEq}
	\ddot{\phi}+\eta\dot{\phi}-\partial_{xx}\phi-\partial_{yy}\phi+\partial_{\phi} V\pap{\phi}=\xi\pap{x,y,t},
\end{equation}
where $\eta$ is a global damping constant, $V(\phi)$ is a generic Ginzburg-Landau potential given by $V\pap{\phi}=\frac{1}{8}\pap{\phi^4 - 2\epsilon\phi^2+1}$ and  $\xi\pap{x,y,t}$ is a real white noise with zero mean. 
This potential has two possible minima given by $\phi_{\rm min}=\pm\sqrt{\epsilon}$. The system undergoes a second-order phase transition as $\epsilon$ changes sign from negative to positive. For simplicity, we linearly quench $\epsilon$ as $\epsilon(t)=t/\tau_Q$, where $\tau_Q$ is the quench rate. Specifically, we quench the system from $t/\tau_Q=0$ to $t/\tau_Q=10$ and then maintain it at $\epsilon_{f}=10$. When the system reaches a state of local equilibrium, the order parameter saturates at the values $\phi=\pm\sqrt{\epsilon_{f}}=\pm\sqrt{10}$. It is not possible to solve the Eq. (\ref{GLEq}) analytically, and we resort to a numerical approach. In the numerics, we set $\eta=1$ and the amplitude of the noise equal to $10^{-3}$. The length of the system is $L_x=L_y=200$, and we use the Fourier decomposition with $201\times201$ grid points for the periodic boundary conditions.  We use the fourth-order Runge-Kutta method with the time step $\triangle t=0.01$ in the time direction.

\begin{figure*}[htbp]
	\raggedright
	\includegraphics[trim=0cm 0cm 0cm 0cm, clip=False, scale=0.46]{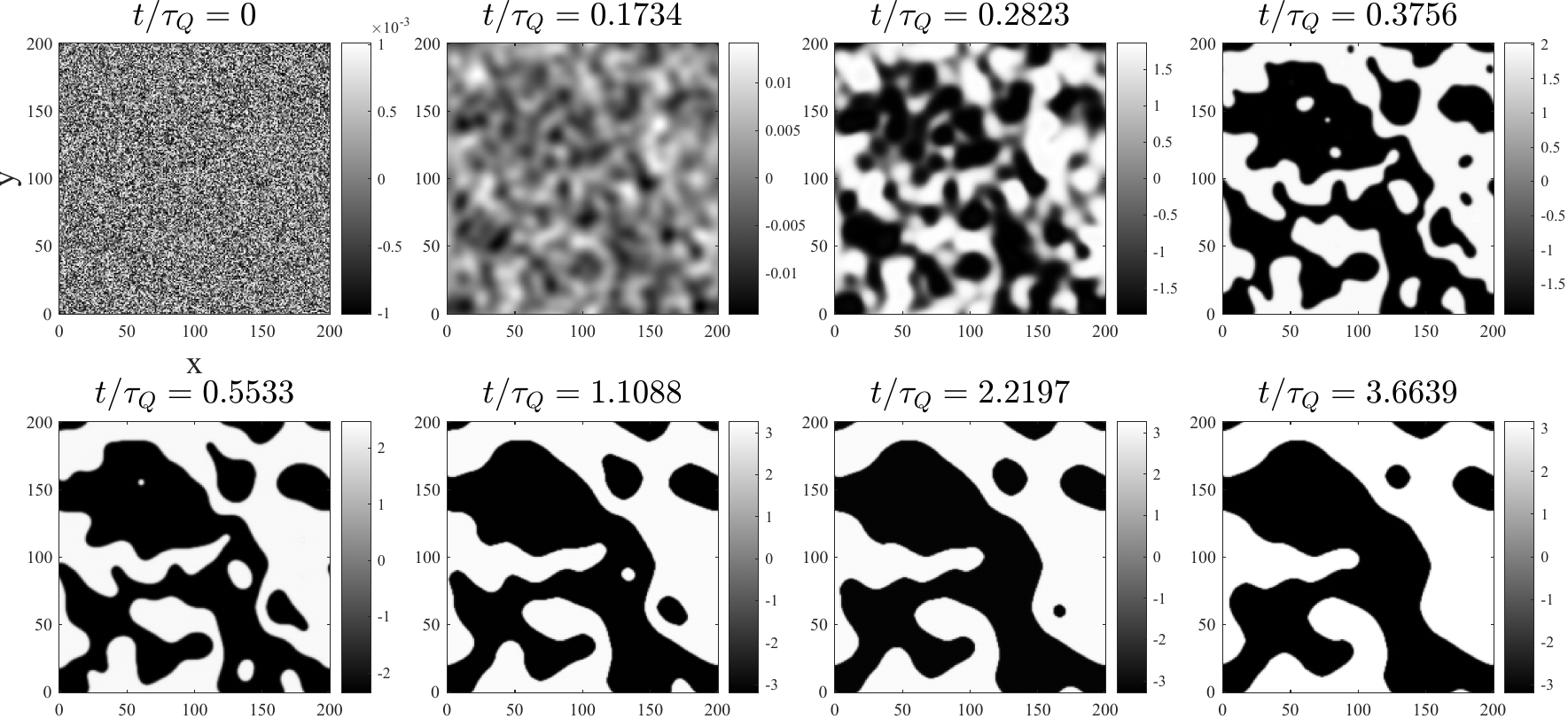}
	\includegraphics[trim=12cm 0cm 0cm 0cm, clip=False, scale=0.53,angle=-90]{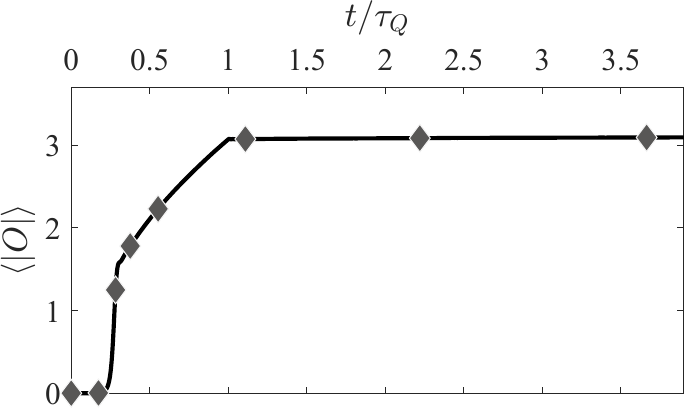}
	\put(-300,-10){\small\bf(a)}
	\put(-65,-10){\small\bf(b)}
	\caption{Time evolution of the order parameter and the birth of domain walls in the TDGL model. {\bf(a)} Density plots of the evolving order parameter at eight specific evolution times 
 with $\tau_Q$ = $e^{4.5}$. One can see how the order parameters evolve from initial random configurations into domain structures. {\bf(b)} The average absolute value of the order parameter $\langle |O|\rangle$ during quenches.  The eight diamond symbols correspond to the eight snapshots in panel (a). }\label{glevolution}
\end{figure*}

Figure \ref{glevolution}(a) shows eight snapshots of the time evolution of the domain profiles of the order parameter. Their corresponding average values $\langle|O|\rangle$ are shown as eight diamonds in the right panel (b) in Fig. \ref{glevolution}. At $t/\tau_Q=0$, the scalar field takes very small random values that serve as the inhomogeneous seeds for the time evolution of the system. At the early time $t/\tau_Q=0.1734$ and $t/\tau_Q=0.2823$, domains formed because of the KZM.  From $t/\tau_Q=0.3756$ to $t/\tau_Q=3.6639$, due to the influence of coarsening, it can be found that the domains merge and form a larger domain. 

\begin{figure}[htbp]
	\centering
	\includegraphics[trim=0cm 0cm 0cm 0cm, clip=true, scale=0.6]{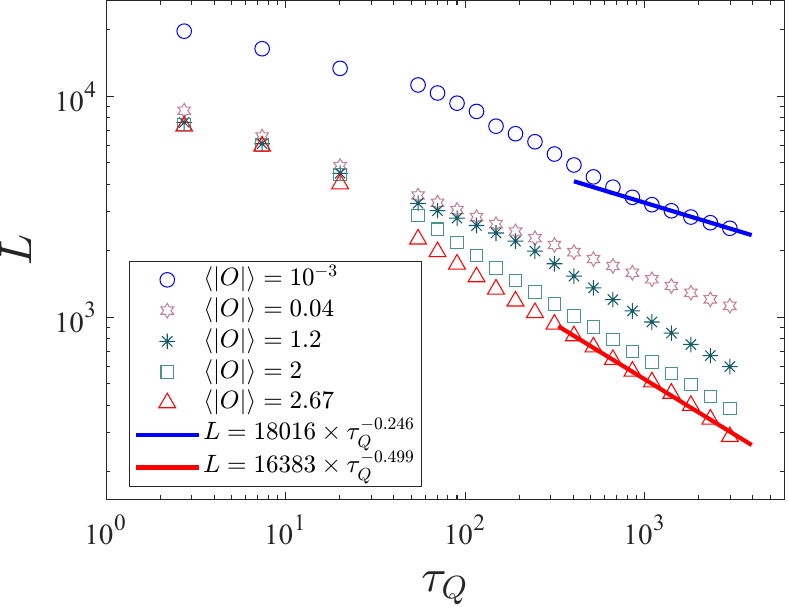}\put(-210,180){(a)}~
	\includegraphics[trim=0cm 0cm 0cm 0cm, clip=true, scale=0.6]{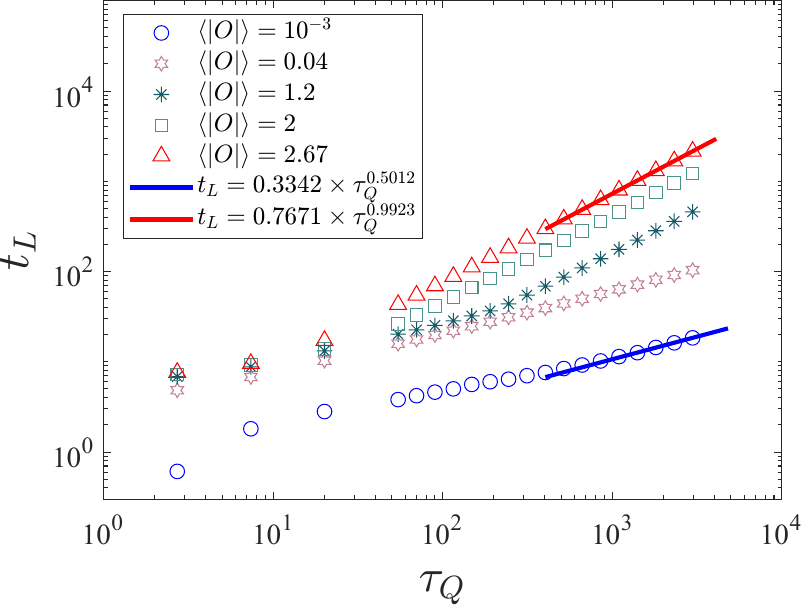}\put(-210,180){(b)}
	\caption{(a) Double logarithmic relation between the domain wall length $L$ and quench rate $\tau_Q$ for the average absolute value of the order parameter $\langle |O|\rangle$.
 (b) The relation between the lag and quench times for various condensates. The circles, hexagons, stars, squares, and triangles are the numerical data, while the straight lines are the best-fitted lines in the slow quench regime (larger $\tau_Q$).}\label{glquench}
\end{figure}

Figure \ref{glquench}(a) displays the length of the domain walls as a function of the quench rate for several values of the average absolute value of the order parameter $\langle |O|\rangle$. As in the holographic case shown in Fig. \ref{hoquench} in the main text, we see that when $\langle|O|\rangle$ is small (e.g.,  for $\langle|O|\rangle=10^{-3}$), the power-law exponent for slow quenches is $-0.246$, which is close to the KZM prediction $-1/4$ in Eq. \eqref{scaling relation} in the main text. However,  for larger condensates (say, when $\langle|O|\rangle=2.67$), the scaling becomes $-0.499$, which rules out the KZM scaling. 
Fig. \ref{glquench}(b) exhibits the dependence of the lag time on the quench rate for various condensates. As in the case in holography (see Fig. \ref{hoquench}(b)), for small condensates and fast quenches, the scaling relation is $t_L\sim\tau_Q^{0.5012}$, which agrees with theoretical predictions Eq. \eqref{t-hat} in the main text. However, as $\langle|O|\rangle$ is large, this relation becomes $L\sim\tau_Q^{0.9923}$, which differs from the KZM prediction. Therefore, as the system is close to the critical point, the scaling relations are consistent with the KZM. However, as the system is driven away from the critical point, the scaling relations deviate from the KZM predictions. We explain this phenomenology in the following. 

\begin{figure}[htbp]
	\centering
	\includegraphics[trim=0cm 0cm 0cm 0cm, clip=true, scale=0.6]{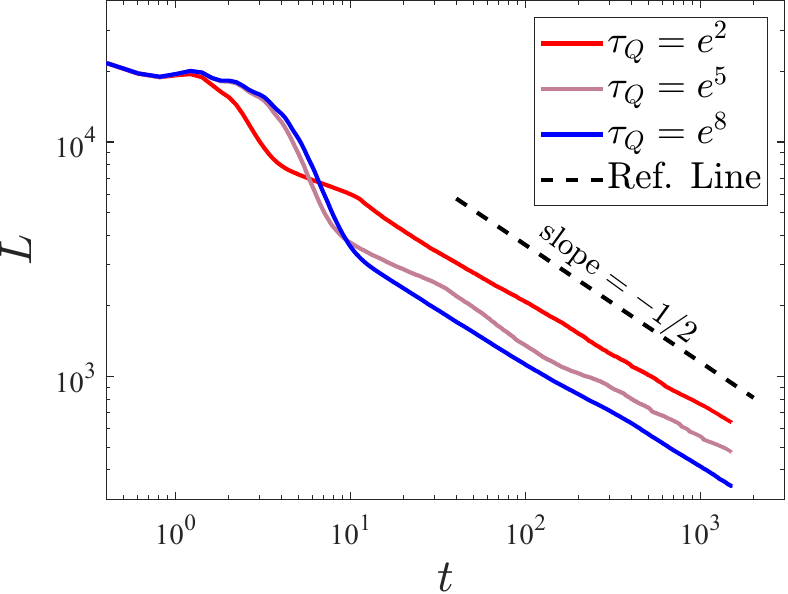}
\put(-210,180){(a)}~~~
 \includegraphics[trim=3cm 9.2cm 3cm 10cm, clip=true, scale=0.6]{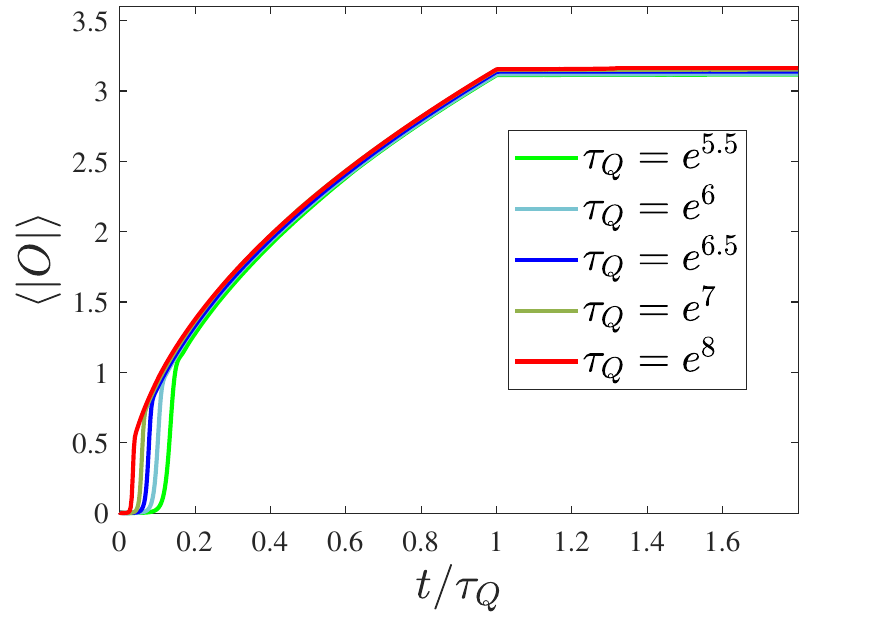}
 \put(-235,180){(b)}
	\caption{(a) Time evolution of domain wall length for various quench rates $\tau_Q =e^{2}, e^{5}$ and $e^{8}$. The black dashed line is the theoretical reference line with a slope of $-1/2$. (b) The average condensate as a function of $t/\tau_Q$ for relatively slow quenches. The overlap at the late stage indicates the adiabatic growth of $\langle|O|\rangle$.}
 \label{gllengthofDW}
\end{figure}

Figure \ref{gllengthofDW}(a) shows the time evolution of domain wall length $L$ for three different values of $\tau_Q$. In the later stage of the evolution, the scaling of $L$ is consistent with the theoretical relation $L\sim t^{-1/2}$, which is similar to the case in holography in the main text. Figure \ref{gllengthofDW}(b) shows the time evolution of the average condensate $\langle|O|\rangle$ for slow quenches. As $\langle|O|\rangle$ is relatively large, the condensates for various quenches overlap due to the adiabatic growth for the slow quench. Therefore, as $\langle|O|\rangle$ is relatively large, such as for $\langle|O|\rangle=2.67$, the overlapping of the lines indicates that the lag-time is linearly proportional to the quench time, i.e., $t_L\sim\tau_Q$, for the slow quenches. This explains the linear relation between $t_L$ and $\tau_Q$ in the Fig. \ref{glquench}(b). Furthermore, from the relations $L\sim t^{-1/2}$ and $t_L\sim\tau_Q$ in the large $\tau_Q$ and large $\langle|O|\rangle$, the  scaling $L\sim\tau_Q^{-1/2}$  explains the numerical results presented in Fig. \ref{glquench}(a). 

\begin{figure}[htbp]
	\centering
	\includegraphics[trim=0cm -0.4cm 0cm 0cm, clip=true, scale=0.6]{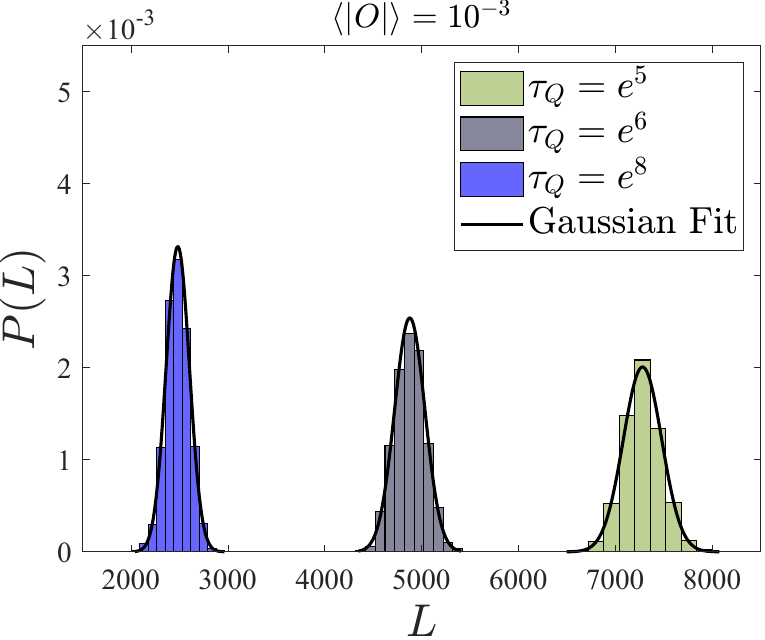}
 \put(-220,180){(a)}~~~
	\includegraphics[trim=0cm -0.4cm 0cm 0cm, clip=true, scale=0.6]{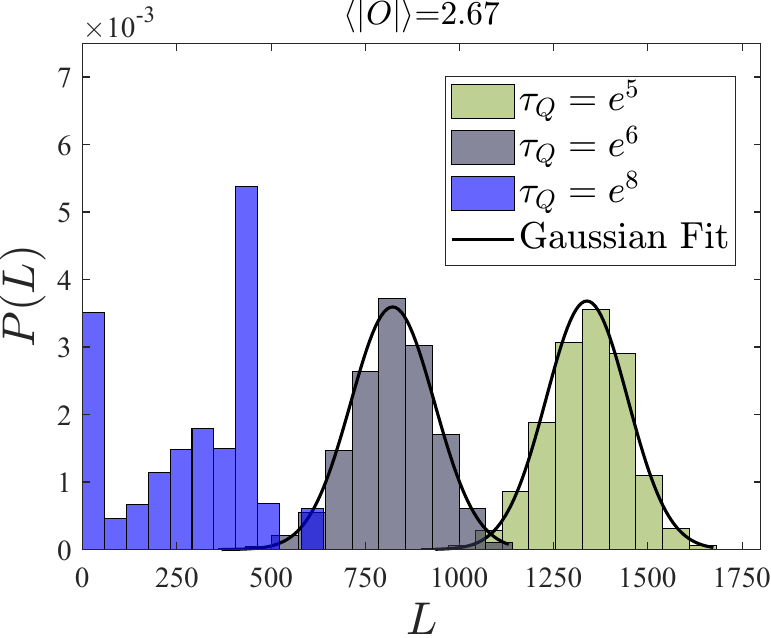}
 \put(-220,180){(b)}~~~
	\caption{(a) Histogram of the domain wall lengths with small condensates $\langle|O|\rangle=10^{-3}$; (b) Histogram of the domain wall lengths with large condensates $\langle|O|\rangle=2.67$. The data involves $2000$  independent simulations. 
 }\label{glhisto}
\end{figure}

Figure \ref{glhisto}(a) exhibits the histogram of the length of the domain wall for various quench rates with small condensates $\langle|O|\rangle=10^{-3}$. For the three values of $\tau_Q$ (i.e., $\tau_Q=e^5, e^6$ and $e^8$), the distributions of the domain wall length with large samplings satisfy the Gaussian distribution. Panel (b) of Fig. \ref{glhisto} shows the histogram of the domain wall lengths with large condensate $\langle|O|\rangle=2.67$. When the quench is not too slow, the distributions satisfy the Gaussian distribution. In the opposite limit (e.g., for $\tau_Q=e^8$, and for a large condensate), the distribution strongly deviates from the Gaussian approximation if the quench is very slow, as coarsening leads to the merging of domains. As a result, events with no domain walls at all become likely, resulting in a multimodal distribution.

\end{document}